\PassOptionsToPackage{hyphens}{url}
\documentclass[letterpaper,twocolumn,10pt]{article}
\usepackage{usenix_2020_09}
\usepackage{epsfig,endnotes}
\usepackage{paralist}
\usepackage{enumerate}
\usepackage{graphicx}
\usepackage{color}
\usepackage{url}
\usepackage{booktabs}
\usepackage{amsmath}
\usepackage{amssymb}
\usepackage{mathtools}
\usepackage{multicol}
\usepackage{balance}
\usepackage{caption}
\usepackage{bigstrut}
\usepackage{xcolor,colortbl}
\usepackage{mathtools, nccmath}
\usepackage{listings}
\usepackage[utf8]{inputenc}
\usepackage{enumitem}
\usepackage{tcolorbox}
\usepackage{algorithm2e}

\begin{document}
\sloppy

\title{Keep your Identity Small: Privacy-preserving Client-side Fingerprinting}

\author{
{\rm Alberto Fernandez-de-Retana}\\
University of Deusto 
\and
{\rm Igor Santos-Grueiro}\\
HP Labs
} 

\maketitle

\begin{abstract}
Device fingerprinting is a widely used technique that allows a third party to
identify a particular device. Applications of device fingerprinting include
authentication, attacker identification, or software license binding.

Device fingerprinting is also used on the web as a method for
identifying users. Unfortunately, one of its most widespread uses is to
identify users visiting different websites and thus build their browsing
history. This constitutes a specific type of web tracking that poses a threat to
users' privacy. While many anti-tracking solutions have been proposed, all of
them block or tamper with device fingerprinting techniques rather than just blocking
their web tracking application. Therefore, users may be limited in their
experience while using a website.

In this paper, we propose \textit{Privacy-preserving Client-side Fingerprinting}
(PCF), a new method that allows device fingerprinting on the web, while blocks
the possibility of performing web tracking. To this end, PCF is built upon
fingerprinting transparency: any website ought to declare its fingerprinting
scripts while users will compute them in a privacy-preserving manner, limiting
the resultant fingerprints for each different domain and, therefore, making
web tracking not feasible.
\end{abstract}

\section{Introduction}
\label{sec:intro}

Device fingerprinting is a common practice on the Internet to uniquely identify
a user~\cite{sanchezrola-dimva2018-knock}.  Original applications of device
fingerprinting include different tasks such as network
identification~\cite{franklin2006passive,banerjee2011wireless}, device
authentication~\cite{boning1996statistical,bowman2002impact,gassend2002silicon,lee2004technique,pappu2002physical},
or attackers detection~\cite{kohno2005remote, fink2007statistical}.

While these applications are still used, device fingerprinting is associated
with web tracking, since it allows third-party agents to uniquely identify
browsers and compile users' browsing history. This application is used for many
companies in conjunction with cookies to track devices for different purposes
such as targeted advertising, fraud detection, content personalization among
others~\cite{acar2013fpdetective, acar2014web, englehardt2016online,
lerner2016internet,sanchezrola-dimva2018-knock,dambrasally}.

\vspace*{0.3em}

Several anti-fingerprinting solutions have been proposed: blacklisting browser
extensions~\cite{ghostery, disconnect} and fingerprint \textit{spoofing
techniques}~\cite{nikiforakis2015privaricator,torres2015fp,laperdrix2017fprandom}.
Despite the fact that these techniques effectively solve web tracking
techniques, they block any fingerprinting method regardless of its nature.  Web
tracking poses a serious threat to users’ privacy and anonymity. However,
device fingerprinting techniques can be used to enhance user experience,
security, or content. Hence, blocking them without any more consideration, may
result in losing the actual functionality of the visited website.

While existing anti-fingerprinting solutions block or limit the device
fingerprinting techniques with little or no consideration of the possible
impact for users' experience on the web, Torres et. al.~\cite{torres2015fp}
tried to tackle this problem with \textit{web identities}, generating a
fingerprint for each website. However, their approach limited the possible
fingerprinting methods and therefore, it also affected seriously its widespread
application.

\vspace*{0.3em} In this paper, we aim at filling this gap between preserving
users' privacy while maintaining web functionality and consistency. We present
\textit{Privacy-preserving Client-side Fingerprinting} (PCF), a device
fingerprinting protocol that allows device fingerprinting behaviors from hosts,
but preserves the users' privacy, removing the possibility of web tracking when
applied.

This paper presents a key insight by proposing a standardized approach that
permits the legitimate use of device fingerprinting for purposes such as
two-factor authentication (2FA) and bot detection, while restricting tracking
solely to a per-domain basis. In our approach, websites will declare the
scripts containing fingerprinting methods. The client receives the website as
usual, and the identified fingerprint container scripts are executed in
isolation from the rest of the web page. By utilizing Browser WebAPI's, such as
screen resolution, these fingerprinting scripts are granted access to the
actual device features. To mitigate the potential tracking risks associated
with the utilization of real values, these scripts are also isolated in terms
of communication, with only a few exceptional cases allowed for data exchange.
Section \ref{sec:model} provides an explanation of these exceptional cases,
which enable per-domain tracking, bot detection, and two-factor authentication
(2FA).

\subsection{Contributions}

In summary, the main contributions of this paper are:

\begin{itemize}
    \item We propose PCF, the first privacy-preserving client-side fingerprinting protocol.
    \item We demonstrate that our protocol
      preserves client privacy, while allowing legitimate device fingerprinting uses (e.g., bot detection, 2FA, per-domain tracking).
    \item We explain how PCF should be implemented by browser vendors to allow the legitimate use of device fingerprinting by websites.
    \item We provide a detailed description of how websites can effectively employ this protocol for per-domain tracking, bot detection, and two-factor authentication (2FA). 
    \item We developed a standard that can be extended to accommodate future legitimate use cases.
\end{itemize}

\section{Background}
\label{sec:background}

\subsection{Fingerprinting Actors}
Online tracking and device fingerprinting is a very complex ecosystem with
several involved parties that play several roles. We provide a summarization of
the most relevant party roles that apply to PCF.

\begin{itemize}
\item \textbf{Client.} The client is the agent that visits a host and is a
  potential target to be fingerprinted (willing or unwillingly). In general, we
  can consider the client as the set of possible fingerprints that a
  fingerprinting script provider may use. This includes the browser, and every
  feature accessible from a client-side script (e.g., browsing data, list of
  installed fonts, and so on).
\item \textbf{Fingerprinting Script Provider.} This party is the one that
  instructs the final client to execute a fingerprinting script. This party can
  be the host that the particular client is visiting (and, therefore, she may
  be aware of its existence) or can it be loaded as part of a third-party site as
  an iframe, an external import, and so on.
\item \textbf{Fingerprint User.} Once the fingerprint is generated in the client
  by the provided script, the resultant ID is sent to a certain party. That
  party can be a script provider, storing the clients' browsing history, or be
  another third-party. In addition, since stateless fingerprints are not stored
  in the client, parties that receive the client's fingerprint may be able to
  share it with other parties, increasing the impact of a potential web tracking
  on that particular client. In addition to receiving just the ID, it is possible 
  for this actor to also receive additional information about the client, such as,
  the screen resolution, fonts or the language.
\item \textbf{Visited Host.} This party is the one which the client
  willingly decides to visit. Currently, a website includes much
  third-party content (imported or directly rendered) that may or may not be a
  potential \textit{fingerprinting script provider} or \textit{fingerprint
    user}, and the visited host may also act as one, or both categories.
\end{itemize}

\subsection{Fingerprinting Techniques}

Fingerprinting methods' objective is to uniquely identify a target. This
identification can be used for a variety of applications including device
authentication, software license binding~\cite{fink2007statistical,sanchez18clock}, wireless network
identification~\cite{banerjee2011wireless, franklin2006passive}, or attackers
identification~\cite{kohno2005remote, fink2007statistical}.

In the particular website of fingerprinting in the web scenario, a website owner
(or a third-party linked by this website) computes a unique fingerprint for each
user, without any storage on the client-side. This is the reason why these
approaches are called stateless and are hard to block --- when they are used for
web tracking.  Sanchez, Santos, and Balzarotti~\cite{sanchez18clock} classified
device fingerprinting techniques into two main groups depending on the features
used:

\begin{itemize}

\item \textbf{Attribute-based Fingerprinting:} These fingerprinting techniques
  use attributes accessible through the browser e.g., browser attributes such as the
  list of installed fonts, the \texttt{UserAgent}, or the screen resolution.
  Their main advantage is that their computing overhead is small, and they are
  easy to obtain. However, the features used in these fingerprinting techniques
  might be changed by users during an update, and therefore, they are considered
  more ephemeral (for a comprehensive study of their ephemeral nature, please
  refer to~\cite{fpstalker}).
    Many attribute-based methods exist for fingerprinting such as fonts
\cite{eckersley2010unique}, or the combination of browser attributes
\cite{laperdrix2016beauty}.

\item \textbf{Hardware-based Fingerprinting:} In order to avoid the problems of
  attribute-based fingerprinting, hardware-level features have been used to
  create a more precise form of fingerprinting. These methods employ differences
  in the hardware that are subtly detectable by calling certain APIs that use
  the underlying hardware in order to compute differences amid
  devices. \textit{Canvas Fingerprint} and \textit{WebGL}~\cite{mowery2012pixel}
  compute the subtle differences in the form text is rendered by the HTML5
  Canvas or WebGL. Another notable web hardware-based fingerprinting technique
  \textit{CryptoFP}~\cite{sanchez18clock} derives the quartz clock manufacturing
  differences, by exploiting the usage of fast Cryptography API functions.

\end{itemize}

\subsection{Application to Web Tracking}

The relation between web tracking and device fingerprinting is old. Web tracking
is a very common technique used on the Internet to retrieve user browsing data
initially  introduced for web advertisement or
analytics~\cite{lerner2016internet}.  This practice has changed its original
goal over the years becoming a widely popular method for a variety of different
goals.

Since web tracking allows third parties to discover users' browsing
history, it can be used for improving users' experience and browsing. However,
because they involve gathering users' data, web tracking can be considered
``invasive'' --- at the very least~\cite{sanchez2016web}.

Many techniques exist that allow web tracking. The first one introduced was the
\textit{cookies}~\cite{soltani2010flash} that, even though its was not its
primary goal when they were designed, it is still considered a popular
technique for tracking (including users' cookie sharing and other privacy
problems~\cite{ssp21_cookieflows,iqbal2022khaleesi,patentbouncetracking}).
Evercookies, to bypass cookie cleaning, cookie syncing to allow trackers to
share users' fingerprints, or ETags inserted within images used to check the
user identity~\cite{ayenson2011flash}; are some examples of web tracking early
evolution.  Device fingerprinting techniques create a unique id for the user
without storing it in the machine (such as installed components such as fonts
\cite{eckersley2010unique}, the computed difference in rendering text by the
HTML5 Canvas API \cite{mowery2012pixel}, the combination of browser attributes
\cite{laperdrix2016beauty}, or timing difference when CPU clock is
stressed~\cite{sanchez18clock}).

In this way, web tracking techniques have been classified regarding their need
of storage~\cite{sanchez2016web,sanchez2015tracking}.  Stateful techniques (such
as cookies) required storing the ids within the client machine, whereas
stateless techniques (such as Canvas or CryptoFP fingerprinting) do not require
that because they are computed each time the user visits a website. Stateless
tracking and fingerprinting techniques are considered more dangerous because
they are harder to limit or block, bypassing common
countermeasures~\cite{acar2013fpdetective}.  Furthermore, a recent work has
measured tracking from users' side (using user telemetry data) and realized
that, if users' general navigation trends are taken into account when measuring
tracking, its prevalence and its impact on users' privacy is at least twice as
estimated before~\cite{dambrasally}.

Web tracking is a very widespread issue on the web.
Studies~\cite{englehardt2016online,sanchezrola-dimva2018-knock} show that more
than 90\% of the websites include (either as first or third party) at least one
script with any sort of tracking behavior. In addition, web tracking is also a
very profitable endeavor that generates billions of
dollars~\cite{iab2020,lau2020brief}. Despite the fact that device fingerprinting
based web tracking is hugely prevalent on the Internet, end-users are not aware
of it or its consequences. Recent works have tried to analyze users' perspective
in this matter and the results showed that users are generally surprised or
willing to adopt more private browsing when they discover the actual data they
are providing~\cite{melicher2016preferences, weinshel2019oh}.

In summary, device fingerprinting (and particular stateless fingerprinting) is
essentially any identification technique, which is difficult to block, and
widely used for web tracking that presents several privacy concerns for
users.

\subsection{Privacy Risk and Threat Model}

Users privacy in device fingerprinting is an issue reported by numerous
studies~\cite{acar2013fpdetective,fpstalker,sanchez2015tracking,lerner2016internet,englehardt2016online,sanchezrola-dimva2018-knock,durey2021fp}
which users usually are not even aware of~\cite{melicher2016preferences,
weinshel2019oh} since no private alternative for device fingerprinting exist.
However, legitimate uses of device fingerprinting exists such as software
license binding~\cite{fink2007statistical,sanchez18clock}, device
authentication, network discovery~\cite{banerjee2011wireless,
franklin2006passive}, or attack
detection~\cite{kohno2005remote,fink2007statistical}.

Imagine the agents involved in fingerprinting: (i) the client, (ii) the
fingerprinting script provider, (iii) the fingerprint user, and (iv) the visited
host. In a typical interaction:

\begin{enumerate}
\item A \textit{client} will visit a host \texttt{www.example.com}. This site
  will be the \textit{visited host}.
\item The \textit{visited host} may include several fingerprinting scripts
  internally or imported as a third party. In our case, we will imagine 3
  domains that use 3 fingerprint scripts, \texttt{fp1.js}, \texttt{fp2.js}, and
  \texttt{fp3.js} from \texttt{a.com}, \texttt{b.com}, and the \textit{visited
    host} \texttt{www.example.com}.  These domains will be considered
  \textit{fingerprinting script providers}.
\item On the client side, the
  scripts \texttt{fp1.js}, \texttt{fp2.js}, and \texttt{fp3.js}; are imported
  and executed generating 3 unique fingerprints: $f1$, $f_2$, and $f_3$. Usually
  these fingerprinting algorithms will compute the fingerprint within the
  client's machine and will send the results back to the original
  \textit{fingerprinting script providers} domains or other domains. Any domain
  that receives a client's fingerprint is considered a \textit{fingerprint
    user}.
\item The different \textit{fingerprint users} will compare these fingerprints
  with the ones already collected to identify the client and act accordingly.
  In some case, it will only imply personalization or even a second factor of
  authentication, while in many cases since fingerprints can be shared with
  other \textit{fingerprint users} or the third-party \textit{fingerprinting
    script provider} may include their fingerprinting script in many sites,
  fingerprints are used to retrieve browsing history performing web tracking.
\end{enumerate}

This problem of web tracking and stateless device fingerprinting in particular,
is the difficulty to opt-out by the
user~\cite{acar2013fpdetective,sanchez2016web,sanchez18clock}. When we consider
the current legislation and recommendations, the client must be given the choice
to opt-out any type of user identification (e.g., GDPR or California privacy
law). However, these legislations are only followed in the case of third-party
cookies because of its stateful nature --- and even in this case, the
application of the law is not as strict as it should~\cite{sanchez2019can}.

\subsection{Legitimate Device Fingerprinting}
Device fingerprinting, when used responsibly and within appropriate contexts,
can serve several legitimate purposes that contribute to user security,
authentication, and website functionality~\cite{laperdrix2020browser}. While
there is a significant body of research focusing on techniques to block or mask
device fingerprinting for privacy protection, there is a noticeable gap in
literature when it comes to exploring a middle path that allows for the
legitimate use of device fingerprinting.

One such purpose is user
authentication~\cite{furkan2016device,nampoina2021authentication,andriamilanto2020fpselect,van2016accelerometer},
where device fingerprinting can be utilized to verify the identity of a user
based on unique device characteristics. This technique could be used as a
two-factor authentication (2FA). By analyzing factors such as browser
configuration, installed plugins, and system attributes, websites can develop
mechanisms that aids in distinguishing legitimate users from potential
impersonators or unauthorized access attempts. Another legitimate purpose for
device fingerprinting is bot
detection~\cite{jonker2019fingerprint,vastel2020fp,xigao2021good}. With the
help of device fingerprinting, websites can analyze various device
characteristics, behavior patterns, and fingerprinting attributes to
differentiate between legitimate users and automated bots or malicious scripts.
By implementing robust bot detection mechanisms based on device fingerprinting,
websites can effectively mitigate fraudulent activities and maintain the
integrity of their services.  In addition to the aforementioned purposes,
device fingerprinting can also be utilized to determine if the client software,
including the operating system or browser, is outdated. This information
enables websites to prompt users to update their software for improved security
and compatibility, or restrict access to certain services until the necessary
updates are applied. By leveraging device fingerprinting for software version
detection, websites can enhance security measures and protect against potential
vulnerabilities.

Finally, another legitimate goal of device fingerprinting is per-domain
tracking, which is distinct from privacy-invasive web tracking. Per-domain
tracking allows websites to track user activity within their own domain for
various purposes, such as personalization, analytics, and security. By
utilizing device fingerprinting techniques, websites can identify returning
users and track their interactions within a specific domain, providing a more
tailored and customized experience. It enables websites to maintain session
information, remember user preferences, and deliver targeted content, all
within the confines of their own domain and without infringing on user privacy
across different websites.

\subsection{Motivations for PCF}

Currently, browser vendors employ various protection and mitigation solutions
to prevent web tracking. These measures range from basic approaches such as
script blacklisting to more sophisticated implementations like randomization or
masking of web APIs. It is important to note that both web tracking scripts and
legitimate device fingerprinting scripts currently operate within the same
execution context. While many publications and implementations address the
blocking or masking of device fingerprinting for privacy concerns, there is a
gap in mechanisms that allow for legitimate use cases of device fingerprinting.
Current approaches may hinder the potential benefits in areas like security,
authentication, and fraud detection. Given this context, we propose the need
for a mechanism that allows the legitimate use of device fingerprinting for
specific purposes, all while prioritizing the protection of user privacy.

In this contribution we present \textit{Privacy-preserving Client-side
Fingerprinting} (PCF) protocol. PCF will allow \textit{fingerprint users} and
\textit{fingerprinting script providers} to utilize the real device features,
overcoming existing blocking mechanisms that are prevalent today in browser
vendors. The primary objective of our protocol is to prevent tracking while
still enabling the legitimate utilization of device fingerprinting for
essential purposes such as security, authentication or per-domain tracking. In
essence, our approach represents a middle ground between blocking
fingerprinting altogether and permitting device fingerprinting without any
restrictions violating users' privacy. 

\subsection*{Adherence to Existing Regulations}

Currently, laws such as the GDPR provide users with the option to opt-out of
certain data collection practices, including device fingerprinting. It is
anticipated that there will be an increasing focus on privacy awareness and the
enactment of additional laws to protect user privacy in the future. These laws
aim to give users more control over their personal data and ensure transparency
and accountability in data processing practices. \textit{Privacy-preserving
Client-side Fingerprinting} (PCF) is designed to enforce privacy-compliant
environments, aligning with the principles of laws like the GDPR. By
implementing PCF, websites can adopt a responsible approach to device
fingerprinting that respects user privacy and provides transparency and control
over data collection practices. PCF enables websites to strike a balance
between legitimate uses of device fingerprinting and protecting user privacy,
ensuring compliance with evolving privacy awareness laws and regulations. In a
similar manner, PCF empowers browser vendors to enforce the blocking of
privacy-harming techniques associated with device fingerprinting in No-PCF
scripts in order to respect existing regulations. 

\subsection*{Reliability}

When adopting PCF, a client should not try to perturb the declared
fingerprints, adding
noise~\cite{nikiforakis2015privaricator,laperdrix2017fprandom} because her
privacy is secured. In a similar vein, the \textit{fingerprinting script
provider} or the \textit{fingerprint users} can trust that the user will not
tamper with the device values or modify them.

\subsection*{Transparency of Used Methods}
One benefit that derives for the protocol itself is that fingerprinting becomes
transparent for client. Since the \textit{fingerprinting script provider} needs
to declare the fingerprinting scripts so they can be executed within the client,
the whole process of fingerprinting becomes more transparent and, therefore,
a lesser shady activity.

\subsection*{Web Tracking}
However, device fingerprinting used for web tracking would not be possible
adhering to PCF policies. Although this is a desired privacy-preserving design
consequence, some \textit{fingerprinting script providers} and
\textit{fingerprint users} would be affected because they use device
fingerprinting to track users.

There may be legitimate reasons to track users outside a domain, such as
advertisement or analytics. To this end, we believe that privacy-preserving
techniques~\cite{toubiana2010adnostic,guha2011privad,fredrikson2011repriv,backes2012obliviad,akkus2012non}
should be used.

\begin{figure*}[t!]
  \centering
  \includegraphics[width=0.85\textwidth]{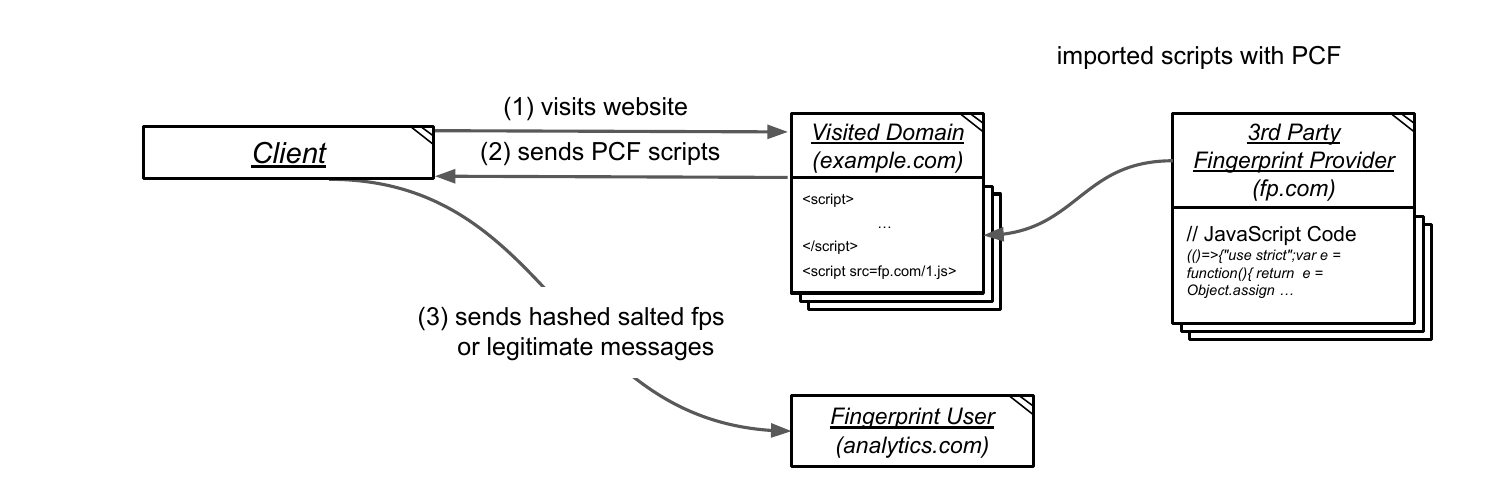}
  \caption{Overview of the \textit{Privacy-preserving Client-side
  Fingerprinting} protocol.}\label{fig:pcf-overview}
\end{figure*}

\begin{algorithm}[t!]
  \footnotesize
    \SetAlgoLined 
    \KwIn{$script$, script included in a website, e.g., the script of a \textit{fingerprinting script provider} .} 
    \tcc{Check if the script is declared as PCF.}
    \eIf{$\neg$ \texttt{markedAsPCF($script$)}}{
      \textbf{normalRuntimeExecution($script$)};
    }{ 
    
      \tcc{Declare allowed communications.} 
      $comms \leftarrow$ \texttt{GenerateAllowedCommunications}()\; 
        
      \tcc{Execute the $script$ in a runtime with the real values of the device and using the declared allowed communications for blocking information exfiltration.} 
      \textbf{PCFRuntimeExecution($script$, $comms$)};
    }
    \vspace*{0.65em}
    \caption{PCF Client-side Algorithm.}\label{alg:pcfclient}
\end{algorithm}

\section{PCF: Privacy-preserving Client-side Fingerprinting}
\label{sec:model}

\subsection{General Overview}
\label{ss:overview}

Device fingerprinting's intended use was originally to provide an identification
method for a particular device or client. However, its usage for unwanted web
tracking is vast~\cite{englehardt2016online,sanchezrola-dimva2018-knock}.

Nevertheless, third parties and websites may want to use these
techniques for legitimate uses. Since current anti-tracking solutions mask or
block these scripts~\cite{nikiforakis2015privaricator,laperdrix2017fprandom}, we
propose that device fingerprinting techniques should not be blocked when used in
a legitimate, transparent manner. Especially when used without compromising user
privacy. Therefore, we present \textit{Privacy-preserving Client-side
Fingerprint} (PCF) protocol that guarantees to a great extent that the computed
users' fingerprints will remain hidden and will not be shared for web tracking,
while their utility for fingerprint providers/users will remain.

The overall behavior within  PCF is as follows (see
Figure~\ref{fig:pcf-overview} for a visual depiction):

\begin{enumerate}
  \item \textit{Client} visits a website using PCF.
  \item \textit{Website}, has its own and third-party
  fingerprinting scripts properly declared as PCF. This declaration implies that each
  script that needs to use the real value of a fingerprinting method, has been declared as PCF.
  Then, \textit{Website} sends all its content, as well as the
  fingerprinting scripts.

  \item \textit{Client} retrieves the content and before executing the scripts, check 
    if any of them is marked with the PCF flag.
  \begin{itemize}
    \item[\textbf{Y}] The script declares the PCF flag, indicating that it
    will execute in an isolated environment where web APIs provide genuine
    values, but communication is limited to declared legitimate messages.
    \item[\textbf{N}] If PCF is not declared, the script will execute in the
    normal mode where browser vendors implement security and privacy measures by
    default to prevent tracking blocking, randomizing and masking specific browser 
    web APIs. 
  \end{itemize}
\end{enumerate}

Scripts imported or loaded from third parties also require to be declared if
they intent to use fingerprinting techniques for legitimate uses. Therefore, it
is the responsibility of each script provider, regardless of whether they are a
first-party or third-party entity, to appropriately indicate the presence of
the PCF flag when it is deemed necessary.  

\subsection{Client-side Fingerprint}
\label{ss:pcf-client}

The core of the computation required for PCF to work is performed on the
client-side. In this way, the client retains control over its fingerprints and
their delivery.

To this end, the client follows the next procedure. When the client visits a
website, the website sends the contents as usual. In the case the website is
adhering to PCF protocol, the scripts that computes device fingerprinting will
be marked with the PCF flag. This includes third-party scripts that the website
may have loaded.

When the script, regardless of whether it is a first-party or third-party
script, declares the PCF header, it operates within an isolated context. Within
this isolated context, the script can access authentic device features.
However, certain restrictions are imposed on communications that extend beyond
this isolated environment to avoid misuses of the protocol. To clarify, when a
script is marked as PCF, it operates within a distinct and isolated runtime
context. Within this PCF runtime context, the script has the capability to
request the genuine device values from the browser, enabling it to accomplish
objectives such as bot detection. However, to prevent any potential web
tracking within the PCF runtime, any communications originating from the script
to other parts of the page or external servers must be subject to appropriate
filtering measures.

This is the general behavior designed for a client in the PCF schema. 
In Section~\ref{sec:design} of the paper, we present a detailed design of the protocol, outlining its various components and considerations. In Section~\ref{sec:diss} of the paper, we delve into a thorough discussion of the most intriguing aspects of the protocol and address the limitations of our contribution. 

\subsection{Fingerprinting Provider}
\label{ss:pcf-fpprovider}

The implementation of the PCF standard is of great interest to \textit{fingerprinting
providers}, as they stand to benefit from its adoption. Fortunately, the
adoption process for \textit{fingerprinting providers} is not overly complex, making it
relatively straightforward for them to incorporate the PCF protocol into their
existing systems. Since the \textit{clients} are the ones that generate the execution
context and communicates back to the \textit{fingerprint user}, the major responsibility
of the \textit{fingerprinting providers} is to provide the means to that to happen. In
other words, within the PCF protocol, the different \textit{fingerprint providers} just
need to declare every script that requires device fingerprinting to allow the
\textit{client} to manage them as described in Section~\ref{ss:pcf-client}.

The required declaration from \textit{fingerprinting providers} will be placed
as HTTP header or as \textit{`script' attribute}. In the next
section~\ref{sec:design}, we will provide a detailed exploration of the
declaration process for PCF scripts. The primary modification that
fingerprinting providers need to make is to encapsulate all device
fingerprinting logic within a single script, allowing it to compute the desired
outcome for the legitimate purpose. Additionally, they would adhere to the
permissible communications outlined in the PCF standard, while being aware that
any other communication would be blocked.

While it is not guaranteed that every \textit{website} or
\textit{fingerprinting provider} will declare all the fingerprinting scripts as
PCF-compliant, the objective of this work is to establish a protocol that
instills confidence in users and websites when executing device fingerprinting.
The implementation of the PCF standard would enable browser vendors to employ
more aggressive mitigations to prevent web tracking in scripts that are not
compliant with PCF, ensuring stronger protection against unauthorized tracking
practices. In this way, we acknowledge and recognize the work of other
researchers in presenting technologies that focus on blocking and mitigating
device fingerprinting (e.g.,
\cite{nikiforakis2015privaricator,laperdrix2017fprandom}), which are fully
complementary to the PCF standard.

\section{Practical adoption in the real world}
\label{sec:design}

Our contribution introduces a protocol that safeguards users privacy whilst allowing the legitimate utilization of device fingerprints,
such as, bot detection or per-domain tracking. In Section~\ref{sec:background},
we provided an overview of the motivations behind introducing a new standard
for device fingerprinting. Section~\ref{sec:model} outlined the high-level
protocol, providing a general understanding of its components. However, in this
section, we delve into the detailed design of the protocol, discussing its
implementation in the real world and providing a comprehensive explanation of its various aspects.

\subsection{PCF Script Declaration}

\begin{figure}[t!]
  \centering
  \includegraphics[width=0.48\textwidth]{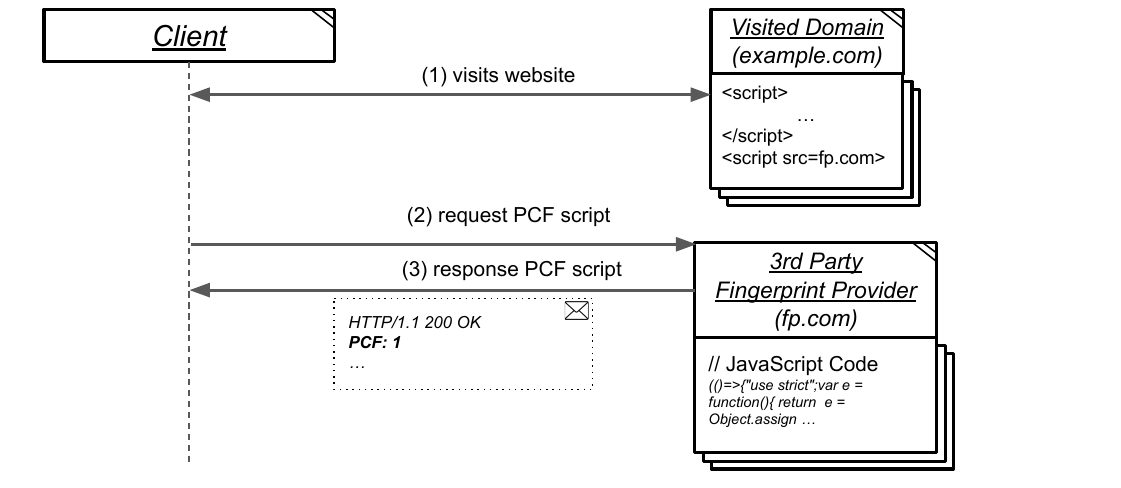}
  \caption{Declaration of \textit{PCF} script by HTTP Header.}\label{fig:pcf-header}
\end{figure}

Web communications provide a large range of opportunities to declare security
or privacy policies, offering a diverse range of possibilities. In the
following lines, we outline our proposed protocol declaration. The initial
method we propose for declaring a PCF script involves implementing a new HTTP
Response Header named after the protocol itself, PCF.
Figure~\ref{fig:pcf-header} provides an illustrative example of this solution. 

In fact, the proposed implementation of the PCF protocol as an HTTP header
response offers a simple and straightforward approach that can be adopted by
both browser vendors and script providers. By integrating the PCF header
response into the HTTP protocol, browser vendors can implement the necessary
mechanisms to enforce the execution context and communication restrictions
defined by the PCF protocol. Script providers, on the other hand, can easily
declare their scripts as PCF-compliant by including the PCF header in the HTTP
response. This solution is similar to other mechanisms implemented in the Web
Ecosystem, such as, \textit{Content-Security-Policy (CSP)},
\textit{X-Frame-Options} or the deprecated \textit{X-XSS-Protection}. However,
this implementation has two main disadvantages. Firstly, in-line scripts cannot be
declared as PCF compliant. Secondly, the first-party is unable to declare PCF by
itself for scripts that are requested to third-parties. 

\begin{figure}[h!]
\footnotesize
\begin{tcolorbox}
\begin{verbatim}
<html>
  <head>
    <script pcf
    src="https://third-party.com/fp.js">
    </script>
  </head>
  // ...
  // Page content
  // ...
</html>
\end{verbatim}
\end{tcolorbox}
\caption{Declaration of PCF scripts as attribute.}\label{fig:pcf-inline}
\end{figure}

To address these limitations, as illustrated in Figure~\ref{fig:pcf-inline}, we
propose an addition to this solution by advocating for the standardization of a
new attribute for declaring scripts as PCF scripts. This proposal
enables first-parties to declare third-party scripts as PCF, allowing their
execution within the new isolated context. By implementing this approach and
ensuring that first-parties have a properly declared Content-Security-Policy
(CSP) header, they acquire enhanced control over the behavior of third-party
scripts. This solution shares similarities with other existing approaches, such
as the \texttt{`sandbox'} attribute in \texttt{`iframe'} elements or the use of \texttt{`nonces'} in \texttt{`script'}
tags.

Our proposal, comprising two implementation designs, offers versatility and ease
in declaring scripts as PCF. The declaration of PCF scripts follows similar
specifications found in the web ecosystem, such as the Permission-Policy, which
encompasses both the HTTP Response Header and the attribute for the
\texttt{`iframe'} tag. Moreover, this compound solution seamlessly integrates
with existing security and privacy headers and attributes without any
compatibility issues. Finally, we would like to emphasize that the
attribute-based solution empowers websites to declare third-party scripts as
PCF-compliant, granting first-parties control over the execution of these
scripts.

\subsection*{PCF Scripts Execution Context}

\subsubsection{WebAPI Execution}
\begin{figure*}[t!]
  \centering
  \includegraphics[width=0.85\textwidth]{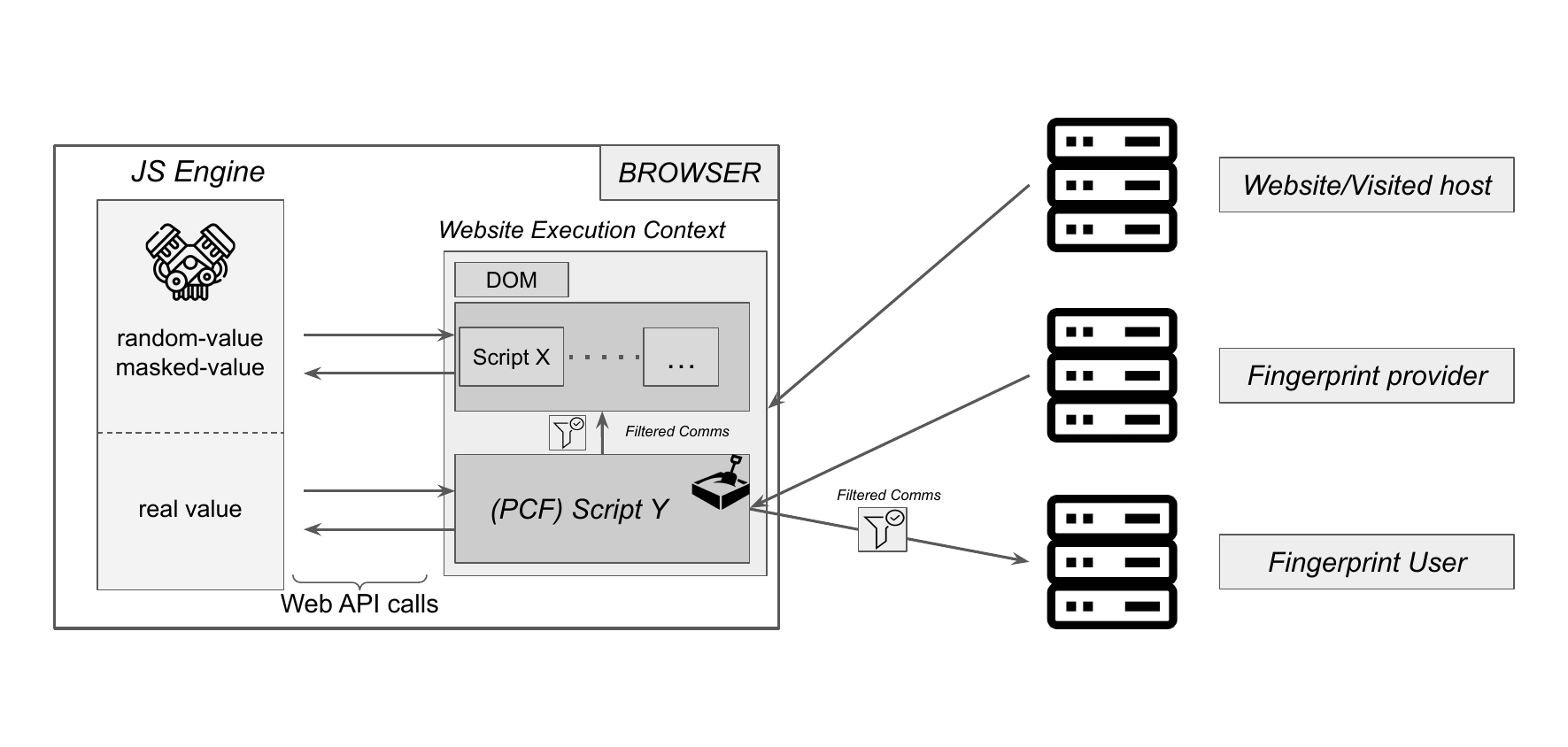}
  \caption{Execution Context of \textit{PCF} implementation.}\label{fig:pcf-execution-context}
\end{figure*}

As elucidated throughout the paper, both web tracking scripts and legitimate
device fingerprinting scripts currently operate within the same execution
context. In this execution context, web APIs are randomized or masked to prevent
web tracking, potentially impacting the legitimate use of these APIs. In this
section we describe the new execution context where declared PCF scripts (as
explained in the previous section) operate within a parallel execution
environment.

Within the PCF Execution context, web APIs would provide authentic values
retrieved from the underlying system. For instance, if the user has a rare UTC
timezone such as UTC-1, browsers, to safeguard the user's privacy, may return a
more commonly used timezone, such as, UTC+2. This privacy-preservation method
fundamentally involves aligning your fingerprint with that of other users,
rendering individual identification unfeasible. However, within the PCF
context, the web API would return the appropriate value, which in this case
would be UTC-1 (real value). By utilizing the actual values provided by various
APIs, the script can make informed determinations, such as distinguishing
between a bot and a legitimate user or identifying if the user is attempting to
log-in from a new device on the webpage. Figure \ref{fig:pcf-execution-context}
illustrates the distinctions between the two execution contexts, highlighting
the variations and implications that arise when executing scripts within the
PCF Execution context compared to the normal execution context.

PCF scripts run in a separate execution context parallel to that of other
scripts on the page. This parallel instance ensures that PCF scripts operate
independently and do not interfere with the execution or behavior of other
scripts running simultaneously. By isolating PCF scripts in their own execution
context, the protocol establishes a clear boundary between PCF operations and
the rest of the script environment, blocking the exfiltration of device
information. In this separate execution context, mechanisms that aim to block
or mask device fingerprinting, such as Brave's
\textit{farbling}~\cite{braveshield}, would be deactivated. This allows PCF
scripts to access and utilize the real device values without any interference
or obfuscation to complete their goal.

\subsubsection*{Script Communications}

Within the PCF framework, we introduce a novel execution context where scripts
have the capability to access the genuine device values. However, this new
context presents a potential risk of unintended exfiltration of these real
device values beyond the PCF script, such as by other scripts or by a external
fingerprinting user. To secure user privacy while enabling the utilization
of device features within the script, we propose implementing filtered
communications from PCF scripts. This approach ensures that only authorized and
necessary data exchanges occur, thereby mitigating the risk of unintended
information leakage.

The implementation of communication blocking would encompass any form of
outgoing communication from the PCF script to external agents, whether it
involves other scripts within the same webpage context or third-party servers.
When it comes to communication with other scripts, there are specific
techniques that should be blocked by-default within the PCF implementation.
Here is a complete list of techniques which needs to be blocked:

\begin{itemize}
  \item \textbf{Global Scope:} If the scripts are defined in the global scope, they can directly access and modify variables and functions defined by other scripts. They can share data by assigning values to global variables or by calling shared functions.
  \item \textbf{DOM Manipulation:} Scripts can interact with the Document Object Model (DOM) to communicate with other scripts. They can access and modify DOM elements, attributes, and properties, allowing them to exchange data and trigger events that can be listened to by other scripts.
  \item \textbf{Event System}: Scripts can utilize an event system to communicate with each other. They can define and dispatch custom events using the \texttt{CustomEvent} API and listen for those events using event listeners. Other scripts can listen for these events and respond accordingly.
  \item \textbf{Shared Storage:} If the scripts need to communicate even when the page is reloaded or reopened, they can utilize shared storage mechanisms such as cookies, local storage, or session storage. Scripts can read from and write to these storage mechanisms to share data between page loads or sessions.
\end{itemize}

The described methods will enable fingerprinting scripts to share genuine
device information with the rest of the webpage. Based on this premise, we
propose the utilization of a partitioned context mechanism for implementing the
set of techniques mentioned above. This approach is similar to the partitioned
context mechanisms currently being implemented by browser
vendors~\cite{storagepartitioning}. It ensures that each script interacts within
a unique scope, storage, or DOM, which is isolated from other scripts and
execution contexts. 

Despite the blocking of all the communications, we now describe the techniques
that would be allowed by the PCF protocol. This filter applies to all web APIs
that interact with the broader web ecosystem, including both the context of the
website and external servers. This filter would impact web APIs such as the
\texttt{PostMessage}, \texttt{XMLHttpRequest}, \texttt{Fetch}, and any other web
APIs that facilitate communication between scripts and external entities. The
purpose of this filtering mechanism is to exclusively allow communications that
adhere to the protocol's guidelines, such as those related to legitimate use of
device features (e.g., bot detection). To this end, we propose a
well-defined set of permissible communications that are in line with these
objectives.

In the proposed filtering mechanism, for external communications only one
request is permitted for each specific purpose and site. This means that for
each distinct purpose and each individual site, a single communication will be
allowed. As defined in the HTML Standard~\cite{sitestandard}, a site refers to a
collection of websites served from the same domain and managed by a single
organization (e.g., shop.example.com and coffee.example.com). Furthermore, these
communications would be restricted to the HTTPS protocol to prevent
man-in-the-middle attacks. In the case of communication to the website context,
the filtering mechanism allows only one communication per purpose. So, for
example, if script `X' needs to communicate with the website context and the
backend server to notify them that the user's software is outdated, the script
would utilize the \texttt{PostMessage} API for one-time communication with the
website context. Subsequently, it would employ the \texttt{Fetch} API to
transmit this information to the backend server. This approach ensures that the
PCF script is enforced to respect the privacy of the user by limiting the
communication to the necessary and allowed channels for the specific purpose of
notifying about outdated software.

Finally, we outline the communication messages that are allowed within the PCF
framework. As mentioned earlier, these messages are designed to be highly
restrictive to mitigate the risk of web tracking or data exfiltration. For
scenarios where the purpose is to determine if the user meets certain
requirements, such as bot detection, fraud detection, and two-factor
authentication (2FA), we propose a payload that consists of a boolean value.
This payload can be used to indicate whether the user fulfills the specified
requirements. When it comes to communicating the user's fingerprint, we allow
the use of any string identifier. However, to ensure privacy and restrict
tracking to specific domains, we propose hashing the identifier with a
domain-specific salt. This hashing process would be performed by the execution
context before sending the message, ensuring that the fingerprint is only used
for tracking purposes within that particular domain. This methodology ensures 
that tracking across different websites becomes impossible, thereby preserving 
user privacy. In our proposal, we suggest using JSON as the default format for
sending data, where the key serves as the identifier for the data being
transmitted.

In summary, communication design is one of the most critical and delicate
aspects of PCF design. A bad communication specification could allow user data
exfiltration, not achieving the intended goals of the protocol and compromising
user privacy. Additionally, it is important to note that the legitimate
payloads for communication are not fixed in a monolithic architecture; they can
be adapted and expanded in the future to accommodate new legitimate
communications that adhere to the PCF standard.

\section{Discussion}
\label{sec:diss}

In this paper, we presented \textit{Privacy-preserving Client-side
Fingerprinting} (PCF) a standard proposal to allow legitimate
device fingerprinting purposes, avoiding the risk of web tracking.
Several aspects are topics of discussion regarding the application
of the PCF method. In this section, we discuss the main implications, design
choices, and limitations of our approach.

\subsection*{Protocol Adoption}
One of the key assumptions of PCF is the widespread adoption and acceptance of
the standard by affected parties: browser vendors/clients, websites and
third-parties. However, the development and implementation of PCF would improve
the lives of all actors involved, except for those whose goal is to
compromise user privacy (e.g., web tracking). 

In the case of websites and third parties, the implementation of the PCF
standard would greatly simplify the adoption and implementation of legitimate
device fingerprinting techniques. Currently, these techniques face challenges
due to browser mitigations and restrictions, such as, web api randomization.
However, by adhering to the PCF standard, websites and third parties can
overcome these limitations and leverage device fingerprinting in a responsible
and effective manner, enabling various use cases such as user authentication,
bot detection, and compatibility checks. For example, a bank webpage could
utilize device fingerprinting to check if a user is running outdated software,
which could potentially make them vulnerable to browser vulnerabilities. Based
on this information, the bank could take appropriate actions, such as blocking
access until the user updates their software.

In the case of clients, the implementation of the PCF standard enables browser
vendors to employ more aggressive strategies for identifying and handling
suspicious scripts. By creating an isolated context for device fingerprinting
scripts, it establishes a controlled environment where legitimate device
fingerprinting can occur, while any activities outside this context that are
recognized as suspicious for web tracking can be effectively blocked. In simple
terms, PCF provides a conducive environment for legitimate device
fingerprinting by allowing various actors to perform such activities. This
proposal enables browser vendors to apply stricter controls on scripts that are
identified as engaging in device fingerprinting, as PCF is specifically
designed to address this purpose. In summary, the adoption of the PCF protocol
offers a practical and straightforward solution for browsers and websites
without requiring complex or unconventional implementations.

\subsection*{Implementation design}
In our paper, we introduce PCF, a protocol that incorporates a diverse set of
design policies to address various considerations and ensure effectiveness and
privacy compliance. In the subsequent paragraphs, we delve into a discussion of
alternative approaches that were considered for each design choice,
highlighting their relevance and potential implications.

\subsection*{PCF Declaration}
We suggest that the declaration of PCF scripts can be done through two methods:
using the HTTP Response Header or the script attribute. In these solutions, our
focus has been primarily on developing the ability to declare whether a script
is PCF-compliant or not, without delving into more detailed aspects of the
declaration process. For instance, in addition to simply selecting whether a
script will be executed as PCF, there could be the possibility to declare the
specific goals or objectives of the script. This would allow for a more nuanced
and detailed declaration of the script's intended purpose within the PCF
framework.

\subsubsection*{One-Time Communication policy}
In PCF, we propose the implementation of a one-time policy for communications.
This policy dictates that each PCF script is limited to one communication event
with the webpage (e.g., using postMessage) and one communication event with
each site~\cite{sitestandard}. By limiting the number of communication events,
we mitigate the risk of unauthorized data transmission. For example, if
multiple connections were allowed per origin, it could potentially enable the
exfiltration of data across different subdomains within the same domain.

\subsubsection*{Other Communications Types}
As mentioned earlier, the default communication policy of PCF is to block, but
there is room for additional possibilities that are not covered in this work,
such as proposing and incorporating legitimate device fingerprinting methods
for web personalization, as long as they do not compromise user privacy. This
proposal remains open for developers and security researchers to explore and
contribute further with newlegitimate device fingerprinting goals.

\subsubsection*{Execution Context Isolation}

In our contribution, we assert that there is an isolation between PCF scripts
and the rest of the page, but we propose that within the PCF environment,
scripts should be able to communicate with each other. This design proposal
enables websites to leverage the functionality and methods developed by
third-party scripts within the secure and controlled PCF environment.

\subsubsection*{User preferences}
Another intriguing aspect that could be considered in the future for PCF is the
inclusion of user preferences. The design of PCF enables clients to choose when
they want to undergo device fingerprinting and when they prefer not to. For
instance, if a user wishes to browse a shopping page without being
fingerprinted, they could deactivate PCF specifically for that site.

\subsection*{User Identification Mechanisms}

PCF offers a well-defined user identification method within a domain, based in
the device features. By incorporating per-user salt in the fingerprint
communication, PCF has the potential to not only identify the device but also
the user associated with that device. This additional layer of salt, combined
with the per-domain salt, can contribute to enhanced user identification within
the PCF framework. Indeed, by incorporating per-user salt in the fingerprint
communication within PCF, websites would have the ability to establish
different webpage configurations based on the individual user. This
customization could enable tailored user experiences, personalized content, and
specific settings based on the identified user within the PCF framework.

\subsection*{Limitations}

PCF's client-side design is a significant limitation, as it requires executing
all the logic for different purposes on the client side. This approach
introduces challenges and complexities, similar to those found in the gaming
ecosystem, where the server must provide the necessary logic and information in
the script. Additionally, the client-side nature of PCF opens up the
possibility of script or communication manipulation, although it is worth
noting that this limitation is not unique to PCF and exists in normal script
execution as well.

Another potential limitation of the protocol is its impact on performance.
Implementing a sandboxed runtime inside the browser, even if only resulting in
slight performance overhead, can introduce additional computational and
resource requirements. The isolation and communication filtering layers
introduced by the sandboxing mechanism may require additional processing power
and memory, potentially affecting the overall performance of the system.

\section{Related Work}
\label{sec:rw}

A significant amount of research has been performed in the area of device
fingerprinting. In particular, research has been made both in developing new
fingerprinting
methods~\cite{eckersley2010unique,laperdrix2016beauty,mowery2012pixel,sanchez18clock},
studying their evolution~\cite{fpstalker}, or their prevalence and
relations~\cite{acar2013fpdetective,sanchez2015tracking,lerner2016internet,englehardt2016online,sanchezrola-dimva2018-knock,durey2021fp}.

However, there is little effort from the community on providing
privacy-preserving methods that maintain the fingerprinting functionalities
while developing a privacy-preserving framework for both users and
fingerprinting providers. The community has focused more on the usual
applications of device fingerprinting rather than fingerprinting itself.

\subsection*{Privacy-preserving Web Advertisement \& Analytics}
The most similar approaches to PCF can be found in privacy-preserving
advertisement proposals. These approaches, instead of just blocking
advertisements on the web, propose methods to avoid the private information
leakage from users while maintaining their functionality intact. 

Toubinana et al.~\cite{toubiana2010adnostic} presented \textit{Adnostic} that
enabled targeted advertisement without compromising users' privacy, performing
her behavioral profiling in the user's browser.
\textit{Privad}~\cite{guha2011privad} is a very similar approach for targeted
advertisement that incorporates other actors to the proposed protocol such as
\textit{ad brokers} or \textit{dealers}.
\textit{RePriv}~\cite{fredrikson2011repriv} explorer further browsers'
capabilities by maintaining an inference model within the browser space for each
user. In addition, Backes et al.~\cite{backes2012obliviad} proposed
\textit{ObliviAD}, a provably secure architecture for privacy-preserving online
behavioral advertisement. In this formal and cryptographic solution, there is
no assumption of any trustable third parties. 

Privacy-preserving methodologies have also been proposed for other domains
directly connected to the usage of web analytics. Akkus et al.\cite{akkus2012non}
presented a non-tracking web analytics system that allowed publishers to
directly measure the information, rather than inferring it, via computing these
statistics within the client.

\subsection*{Anti-fingerprinting Solutions}
Most solutions seeking to protect users' privacy in the realm of device
fingerprinting have been focused on breaking the tracking and linkability of
device fingerprints.

Blocking extensions exist that seek to block previously identified
fingerprinting scripts before loaded by the
browser~\cite{acar2014web,englehardt2016online,lerner2016internet,sanchezrola-dimva2018-knock}
(e.g., Ghostery~\cite{ghostery} and Disconnect~\cite{disconnect}).

In a similar vein, Tor browser is a modified Firefox for the Tor network. It
limits the effects of fingerprinting by making them as uniform as
possible~\cite{torbrowser}. This browser spoofs fingerprinting input values,
modifies and/or removes attributes, making it recognizable by third parties and
the generated fingerprints are hard to maintain.  The main limitation of these
extensions and browser is, because of their blacklisting nature, they require
large files of scripts performing fingerprinting to block them. The evolving
nature of the web, makes it difficult to maintain updated scripts
lists~\cite{laperdrix2017fprandom,peter2020blacklists}.

Another method used to mitigate device fingerprint is to tamper with the expected
results of fingerprinting methods. There are many browser extensions that
\textit{spoof} device fingerprinting, but they do not produce consistent
fingerprints and, therefore, user may lose
functionality~\cite{nikiforakis2015privaricator}.
\textit{Privaricator}~\cite{nikiforakis2015privaricator} is an anti
fingerprinting method that generates randomness in the requested fingerprinting,
but at the same time, establishes several randomization policies to avoid losing
consistency. In similar vein, \textit{FP-Random}~~\cite{laperdrix2017fprandom}
introduces randomness to the particular JavaScript Engines that are used to
generate the fingerprints in order to tamper with the generated fingerprinting while
limiting the impact on consistency.

In contrast to our PCF approach, these approaches can block and limit the
effects of tracking, their goal is to make the device fingerprint useless,
omitting any legitimate use that device fingerprinting may have. However, these
approaches can be used as a complement of PCF in order to detect and block
misuses of the protocol by rogue third parties.

\subsection*{Privacy-preserving fingerprinting}
As aforementioned, most of the existing solutions in the literature seek to
block or mitigate device fingerprinting and, in this way, eliminate any
possibilities for web tracking.

FP-Block~~\cite{torres2015fp} implemented a solution based on the separation of
web identities: FP-Block generated an unique fingerprinting for each host and
that was the one used in the communications with that particular host. Our
contribution, in contrast to FPBlock, goes beyond enabling per-domain
fingerprinting and extends to support other legitimate device fingerprinting
purposes. While FPBlock focuses on blocking fingerprinting scripts on a
per-domain basis, our approach, PCF, provides a framework that allows for the
execution of various types of device fingerprinting techniques for legitimate
purposes. Furthermore, the main problem with FP-Block is that fingerprints
result inconsistent, because of an incomplete coverage of methods used for
fingerprinting~\cite{laperdrix2017fprandom}. Christof's solution needs an
updated list of the methods used for fingerprinting, while could lead to a
situation in which there are no more unique web identities. The client
implementation also needed for the solution implies a complex logic which could
reduce the overall performance. PCF, in contrast, can provide as much
fingerprinting diversity as needed, overcoming these issues.

\subsection*{Online Bot Detection \& Trust}

Elie et al.~\cite{elie2016picasso} presented \textit{Picasso}, a lightweight
device fingerprinting method to detect traffic sent by an emulator simulating a
real device. Picasso is based upon HTML5 canvas graphical primitives. The
authors demonstrated that their tool was able to perfectly distinguish between a
real device, such as iPhone running Safari, from a desktop client spoofing the
same configuration. 

In a similar vein, Google announced Trust Tokens~\cite{trusttokens}, a method
against fraud, capable of distinguishing bots from real humans.  Websites issued
cryptographic tokens to users they trust (e.g., reCAPTCHA score). When users
visited a given website, the server could accept the previously generated token as
a proof that the user is not a bot. These tokens, in contrast to cookies, are
not unique for each user. 

In these approaches, their goal is to detect bots using techniques to identify
and validate the integrity of users. While the goal of these approaches is far
from ours, the techniques presented might be adapted as a complement to enhance
trust among parties within our PCF protocol.

\section{Conclusions}
\label{sec:conc}

We introduced \textit{Privacy-preserving Client-side Fingerprinting} (PCF) to
fill the gap between allowing fingerprinting legitimate uses and blocking web
tracking that threats users' privacy. PCF's goal is to advocate for device
fingerprinting transparency: websites willing to use fingerprinting techniques
while preserving users' privacy must declare their fingerprinting-containing
scripts. In this way, we have demonstrated how the implementation of PCF would
enable the execution of device fingerprinting for legitimate purposes. 

In our contribution, we have provided a comprehensive description of the
necessary implementation steps for the standardization of PCF. We have outlined
the specific requirements for both browser vendors and fingerprinting script
providers, highlighting the responsibilities and actions they need to take to
adopt the PCF protocol.

We showed that PCF makes device fingerprinting easier for both parts, allowing a
durable and reliable fingerprint generation and management, an authentication
method for both client and website; and, overall, a more transparent and
efficient device fingerprinting scenario for both users and hosts.

\balance
\bibliographystyle{plain}
\bibliography{paper}
\end{document}